\documentclass[twocolumn]{autart}    

\usepackage{graphicx}          
\usepackage{epstopdf}
\usepackage{float}
\usepackage{natbib}
\usepackage{amssymb,bm}
\usepackage{latexsym}
\usepackage{enumerate}

\usepackage{algorithm}  
\usepackage{algpseudocode}  
\usepackage{mathptmx}
\usepackage{setspace}
\usepackage{color}

\usepackage[fleqn]{amsmath}
\usepackage{amsfonts}
\usepackage{multirow}
\usepackage{caption}
\newenvironment{proof}[1][Proof]{\par\noindent\textbf{#1.} }{\hfill$\blacksquare$\par}

\begin{document}

\begin{frontmatter}

\title{Distributed model predictive control without terminal cost under inexact distributed optimization} 

\author[delft]{Xiaoyu Liu}\ead{x.liu-20@tudelft.nl},    
\author[kth]{Dimos V. Dimarogonas}\ead{dimos@kth.se},               
\author[kth]{Changxin Liu}\ead{changxinl@ecust.edu.cn},  
\author[delft]{Azita Dabiri}\ead{a.dabiri@tudelft.nl},  
\author[delft]{Bart De Schutter}\ead{b.deschutter@tudelft.nl}

\address[delft]{Delft Center for Systems and Control,
Delft University of Technology, The Netherlands}  
\address[kth]{Division of Decision and Control Systems, School of Electrical Engineering and Computer Science, KTH Royal Institute of Technology, Sweden}        


\begin{abstract}                          
This paper presents a novel distributed model predictive control (MPC) formulation without terminal cost and a corresponding distributed synthesis approach for distributed linear discrete-time systems with coupled constraints. The proposed control scheme introduces an explicit stability condition as an additional constraint based on relaxed dynamic programming.  As a result, contrary to other related approaches, system stability with the developed controller does not rely on designing a terminal cost. A distributed synthesis approach is then introduced to handle the stability constraint locally within each local agent. To solve the underlying optimization problem for distributed MPC, a violation-free distributed optimization approach is developed, using constraint tightening to ensure feasibility throughout iterations. A numerical example demonstrates that the proposed distributed MPC approach ensures closed-loop stability for each feasible control sequence, with each agent computing its control input in parallel.  
\end{abstract}

\end{frontmatter}

\section{Introduction}
Distributed control has been applied in various applications, including connected and automated vehicles, transportation networks, and grid networks, where different agents interact with their neighbors through system dynamics and constraints. 
Distributed model predictive control (MPC) is a widely applied control approach for multi-agent systems \citep{maestre2014distributed}.   In distributed MPC, different agents typically communicate with their neighboring agents to achieve the globally defined objective in a distributed manner \citep{grune2017nonlinear,stewart2010cooperative}. 
The main challenges are ensuring the closed-loop stability of the overall system and designing appropriate distributed synthesis approaches such that each agent can compute its control inputs locally \citep{conte2016distributed,wu2024iterative}.

In order to ensure closed-loop stability under distributed MPC, \cite{conte2016distributed} designed a distributed synthesis approach with separable terminal cost. The same terminal cost was also used to guarantee stability with distributed MPC \citep{stewart2010cooperative,kohler2019distributed,wiltz2025parallelized}. However, the standard approaches for stability in distributed MPC rely on separable terminal cost and terminal sets, which may encounter compatibility issues when terminal cost and terminal sets involve all state variables and cannot be easily separated \citep{giselsson2013feasibility}. Relaxed dynamic programming (RDP) can be used to analyze MPC stability without terminal cost \citep{grune2017nonlinear}.  By using RDP,  \cite{giselsson2013feasibility,rostami2023admm} investigated the stability of distributed MPC without terminal cost. Furthermore, an interpretation using RDP for the stabilization of MPC with terminal cost and terminal sets can be found in \cite{grune2017nonlinear}. 


In distributed MPC, distributed optimization is usually applied to solve the resulting optimization problem \citep{stewart2011cooperative,grancharova2023distributed}. However, due to communication and computational power limitations, many real-time applications of distributed optimization terminate their iteration before reaching the optimal solution, resulting in inexact minimization. 
Such early termination may also result in constraint violation issues, as some dual-decomposition-based optimization approaches, e.g., the alternating direction method of multipliers \citep{boyd2011distributed} and the distributed accelerated gradient algorithm \citep{giselsson2013accelerated}, cannot ensure constraint satisfaction during iterations. Ensuring constraint satisfaction during iterations is critical to applying distributed optimization in distributed MPC, especially for safety-critical systems. \cite{mestres2023distributed} considered distributed optimization problems with separable objective functions and constraints to ensure feasibility in an anytime fashion by designing a forward invariant feasible set.  \cite{wu2023distributed} considered distributed resource allocation problems and proposed a distributed feasible method with specified convergence conditions. 





Tightened constraints is a traditional approach to address constraint violations in distributed optimization. \cite{giselsson2013feasibility,rostami2023admm} addressed constraint violation by tightening state constraints, and they developed an iteration algorithm to ensure constraint satisfaction and recursive feasibility. However, distributed optimization requires an undetermined number of iterations to satisfy the stability condition, which may lead to numerical issues due to excessive iterations. 
In \cite{kohler2019distributed}, a modified optimization approach inspired by robust MPC \citep{chisci2001systems} was developed.  For the chosen non-vanishing tolerance, tightened constraints are defined by using the constraint-tightening approach of robust MPC. By introducing the $k$-step support function \citep{conte2013robust} to calculate such tightened constraints, a modified MPC optimization problem was developed, which provides theoretical guarantees for distributed MPC with inexact dual optimization. However, such non-vanishing tolerance may not be suitable for safety-critical systems, where constraint satisfaction must be ensured at all times.
%

In this paper, we focus on distributed linear discrete-time systems with coupled constraints, and the main aim is to develop a parallel distributed MPC approach that can ensure stability and address the constraint violation issues under inexact distributed optimization. The contributions of the paper are as follows.
\begin{enumerate}
    \item A novel distributed model predictive control (MPC) approach without a terminal cost is developed, in which an explicit stability condition is incorporated as an additional constraint based on relaxed dynamic programming.
    \item A distributed synthesis approach is introduced to handle the stability constraint locally within each local agent, where a violation-free distributed optimization approach is developed to solve the underlying optimization problem, and constraint tightening is introduced to ensure feasibility throughout the iterations.
\end{enumerate}

The remainder of the paper is organized as follows. Section~\ref{preliminaries} presents the preliminaries. Section~\ref{section_stability} provides the stability condition for distributed MPC without terminal costs. Section~\ref{section_distributed} presents a distributed synthesis approach for cooperative distributed MPC. Section~\ref{section_numerical} provides a numerical example and Section~\ref{conclusions} concludes the paper.

\section{Preliminaries}\label{preliminaries}
\subsection{Notation}
A continuous function $f(\cdot): \mathbb{R}_+ \rightarrow \mathbb{R}_+$ is  of class ${\mathcal{K}}$, if it is strictly increasing and $f(0)=0$. A continuous function $f(\cdot): \mathbb{R}_+ \rightarrow \mathbb{R}_+$ is of class ${\mathcal{K}}_{\infty}$, if it is of class ${\mathcal{K}}$ and  $ \lim_{u\rightarrow \infty} f(u) = \infty$. The quadratic norm corresponding to a positive definite symmetric matrix $Q$ is defined as  $||x||_{Q}^2 = x^\intercal Q x$.  

Let us consider a distributed system characterized by a graph $\mathcal{G} = ( \mathcal{N}, \mathcal{E} )$ with $\mathcal{N}$ representing the set of nodes (subsystems) and $\mathcal{E}$ being the set of edges. The cardinality of set $\mathcal{N}$ is represented by $|\mathcal{N}|$.   The local state and input of subsystem $i \in \mathcal{N}$ are $x_i \in \mathbb{R}^{n_i}$ and $u_i \in \mathbb{R}^{m_i}$ respectively. A vector consisting of the stacked sub-vectors $x_i$, $i \in \mathcal{N}$ is represented as $\mathrm{col}_{i \in \mathcal{N}}(x_i)$.  The set of neighbors of subsystem $i$ including $i$ itself is defined as $\mathcal{N}_i = \{j | (i,j) \in \mathcal{E} \}$ with $x_{\mathcal{N}_i} = \mathrm{col}_{j \in \mathcal{N}_i}(x_j) \in \mathbb{R}^{n_{\mathcal{N}_i}}$, $n_{\mathcal{N}_i} = \sum_{j\in{\mathcal{N}_i}} n_j$. 

Let us consider inequality constraints indexed by $s \in \mathcal{S}$ with $\mathcal{S}$ being the set of inequality constraints. We define the subgraph $\mathcal{G}^{[s]} = ( \mathcal{N}^{[s]}, \mathcal{E}^{[s]} )$ as the graph induced by the $s$-th inequality constraint, where $\mathcal{N}^{[s]}$ is the set of subsystems affected by constraint $s$, and $\mathcal{E}^{[s]} = \{ (i,j)|(i,j) \in \mathcal{E}, i, j \in \mathcal{N}^{[s]} \} $. 

A set $\mathcal{X}$ is forward invariant for the dynamics $ x(k+1) =Ax(k)+ Bu(k)$ if $\forall x(k) \in \mathcal{X} $, $\exists u(k)$ such that $ x(k+1) \in \mathcal{X} $.

\subsection{Distributed linear discrete-time systems}
The distributed linear system is described as 
\begin{equation}\label{new_system}
    x_{i}(k+1) = A_{i}x_{i}(k) + B_iu_i(k), \ \forall i \in \mathcal{N}.
\end{equation}
where $k$ is the step, $A_i \in \mathbb{R}^{n_i \times n_i}$, and  $B_i \in \mathbb{R}^{n_i \times m_i}$. We consider the states of $i$ to be constrained by inequalities
\begin{align}
 \sum_{ j \in \mathcal{N}_i^{[s]}}g_{sj} (x_{j}) \le b_{s}, \ s \in \mathcal{S}_i,
\end{align}
where $g_{sj} (\cdot)$: $\mathbb{R}^{n_j} \rightarrow \mathbb{R}$ is a continuous convex function, $b_{s} \in \mathbb{R}_{\ge 0}$ is a constant representing the resource to be distributed,  $s$ is the index of inequality constraint, $\mathcal{N}_i^{[s]}$ is the set collecting neighbors of subsystem $i$ coupled through constraint $s$, and $\mathcal{S}_i$ is the set of inequality constraints affecting subsystem $i$.  In particular, if $\mathcal{N}_i^{[s]} = \{ i\}$, then $s$ represents a local constraint for subsystem $i$ and $b_{s} = 0$. 

In this context, the state and input constraints set for subsystem $i$ is 
\begin{align}
\mathcal{X}_{i} := \{&x_{i} \in \mathbb{R}^{n_i}|  \sum_{j \in \mathcal{N}^{[s]}}g_{sj} (x_{j}) \le b_{s} , s \in \mathcal{S}_i \},\\
\mathcal{U}_i = \{& u_i \in \mathbb{R}^{m_i}| u_i \le U_i\},
\end{align}
where $b_i \in \mathbb{R}^{s_i}$ and $l_i \in \mathbb{R}^{m_i}$ are vectors with proper dimension. 

The dynamics of the network can be expressed as
\begin{equation}\label{whole_system}
    x(k+1) = A x(k) + Bu(k). 
\end{equation}
where $x = \mathrm{col}_{i \in \mathcal{N}}(x_i)$, $u= \mathrm{col}_{i \in \mathcal{N}}(u_i)$,  $A\in \mathbb{R}^{n\times n}$, and $B\in \mathbb{R}^{n\times m}$. For compactness, the constraints can be represented by
\begin{align}
&\hspace{-15pt} x \in \mathcal{X} := \mathcal{X}_1 \times \cdots \times \mathcal{X}_{|\mathcal{N}|} \subseteq \mathbb{R}^n,\\
&\hspace{-15pt} u \in \mathcal{U} := \mathcal{U}_1 \times \cdots \times \mathcal{U}_{|\mathcal{N}|} \subseteq \mathbb{R}^m.
\end{align}

{\bf{Assumption 1}} For each $\mathcal{X}_i$, we assume $\{x_{i} \in \mathbb{R}^{n_i}|  g_{si} (x_{i}) \le 0 , s \in \mathcal{S}_i \} \neq \varnothing$.

As $g_{si} (\cdot)$ is a continuous function, if $\inf_{x_i \in \mathcal{X}_i} g_{si} (x_i) > 0$, we can define $\bar g_{si} (x_i) = g_{si} (x_i) - \inf_{x_i \in \mathcal{X}_i} g_{si} (x_i)$ and $\bar b_s = b_s -  \sum_{i \in \mathcal{N}_i^{[s]}}\inf_{x_i \in \mathcal{X}_i} g_{si} (x_i)$. This leads to the new inequality constraint $ \sum_{ j \in \mathcal{N}_i^{[s]}}\bar g_{sj} (x_{j}) \le \bar b_{s}$, with $\bar g_{si} (x_i) \le 0$. In this context, \emph{Assumption 1} still holds. Hence, we can always assume without loss of generality that \emph{Assumption 1} holds.



{\bf{Assumption 2}} (\emph{Viability}) For each $x_i \in \mathcal{X}_i$, $i \in \mathcal{N}$, there exists a $ \mu_i(x_i) \in \mathcal{U}_i$ such that $A_i x_i + B_i\mu_i\big(x_i\big) \in \mathcal{X}_i$ holds.

This assumption defines controlled forward invariance or viability of $\mathcal{X}_i$. It ensures that for all $x_i \in \mathcal{X}_i$ and all $N \in \mathbb{N}_+$, $\mathcal{U}_i(x_i) \neq\varnothing$. \emph{Assumption 2} generally holds and is also stated in \cite{grune2017nonlinear} for nonlinear systems. 



\subsection{Nominal centralized MPC}
Consider the quadratic stage cost of system (\ref{whole_system}) as
\begin{align}
\ell(x,u) = ||x||_{Q}^2 +  ||u||_{R}^2
\end{align}
where 
$Q \in \mathbb{R}^{n\times n}$ and $R \in \mathbb{R}^{m \times m}$ are positive-definite symmetric matrices, and $Q$ and $R$ are block-diagonal matrices of $Q_i \in \mathbb{R}^{n_i\times n_i}$ and $R_i \in \mathbb{R}^{m_i \times m_i}$, respectively. 


The MPC optimization problem for the global network at step $k$ is
\begin{subequations}\label{MPC}
\begin{align}
\hspace{-10pt} \mathop {\min }\limits_{\scriptstyle{\bm{u}}({k})} \  & {J}\left( x({k}),{\bm{u}}({k}) \right) := \sum\limits_{\kappa = {k}}^{k + N - 1} {\ell(x(\kappa|k),u(\kappa|k))}  \label{iMPC_obj}\\
\hspace{-10pt} \mathrm{s.t.}\ \ 
& x({k|k}) = x(k),\\
& x(\kappa+1|k) = Ax(\kappa|k) + Bu(\kappa|k),   \label{MPC_state_dx} \\
& \sum_{ j \in \mathcal{N}^{[s]}}g_{sj} \big(x_{j}(\kappa|k)\big) \le b_{s},\\
& u_i(\kappa|k) \le U_i,\\
&s \in \mathcal{S}, \ i \in \mathcal{N},  \ \kappa = k, \ldots, k + N - 1, \nonumber
\end{align}
\end{subequations}
where $ x({\kappa|k})$ is the predicted state for step $\kappa$ at step $k$  
and $ {\bm{u}}(k) = [u^\intercal(k|k), \ldots, u^\intercal(k+N -1|k) ]^\intercal$. We use ${J}\left( x(k),{\bm{u}}(k) \right)$ to highlight that the objective is a function of the initial value $ x({k})$ and the control sequence ${\bm{u}}({k})$. 


The problem (\ref{MPC}) is a convex optimization problem. Solving (\ref{MPC}) at step $k$ results in the optimal input sequence ${\bm{u}}^{*}({k}) = [u^{*\intercal}(k|k), \ldots, u^{*\intercal}(k+N -1|k) ]^\intercal$; in the moving horizon framework, only the first value $u^{*}(k|k)$ is implemented in the system and the procedure is repeated under a moving horizon scheme.

{\bf{Theorem 1}} \emph{(Recursive Feasibility)} \citep{grune2017nonlinear}. Given that the problem (\ref{MPC}) is feasible at step $k$ with input sequence ${\bm{u}}^{*}({k}) = [u^{*}(k|k), \ldots, u^{*}(k+N -1|k) ]^\intercal$, then, following \emph{Assumption 2},  $\mu\big(x^{*}(k+N|k)\big) \in \mathcal{U}_i$ can be found such that $x(k+N+1|k) \in \mathcal{X}_i$.

{\bf{Theorem 2}} \emph{(Lyapunov Stability)} \citep{grune2017nonlinear}. Considering the system (\ref{whole_system}) with $x(k)\in \mathcal{X}$,  let $\mathcal{X}$ be forward invariant, 
i.e., for all $x(k) \in \mathcal{X}$ there exists $u(k) = \mu(x(k)) \in \mathcal{U}$ such that  $x(k+1) = A x(k) + B\mu\big(x(k)\big) \in \mathcal{X}$.  
If there exist $\alpha \in (0, 1]$, and function $V\big(\cdot \big)$, as well as $\beta_1(\cdot)$, $\beta_2(\cdot)$, and $\beta_3(\cdot) \in {\mathcal{K}}_{\infty}$,  such that for all $ x(k) \in \mathcal{X}$ the following condition (\ref{Stability}) holds, then $V\big(\cdot \big)$ is a Lyapunov function for the system on $\mathcal{X}$ with the equilibrium point $x = \bm{0}$:
\begin{subequations}\label{Stability}
\begin{align}
&\hspace{-20pt} V\big(x({k})\big) \ge \alpha \ell\big(x(k), u(k)\big) + V\big(x({k+1})\big), \label{Stability1}\\
& \hspace{-20pt}\beta_1\big(||x(k) ||_2\big) \le V\big(x({k})\big) \le \beta_2\big(||x(k) ||_2\big), \\
& \hspace{-20pt}\ell\big(x(k), u(k)\big) \ge \beta_3\big(||x(k) ||_2\big). 
\end{align}
\end{subequations}



\section{Stability of Cooperative Distributed MPC without Terminal Cost}\label{section_stability}
In distributed MPC, (\ref{MPC}) is decomposed and the local optimization problem for agent $i$ is
\begin{equation}\label{i_MPC}
 \mathop {\min }\limits_{\scriptstyle{\bm{u}}_i({k})} {J_i}\left( x_i({k}),{\bm{u}}_i({k}) \right)    := \sum\limits_{\kappa = {k}}^{{k} + N - 1} {\ell_i(x_{i}(\kappa|k),u_{i}(\kappa|k))},
\end{equation}
where $\ell_i(x_i,u_i) = ||x_i||_{Q_i}^2 +  ||u_i||_{R_i}^2$. 
The MPC optimization problem of subsystem $i$ is constrained by
\begin{subequations}\label{i_constraints}
\begin{align}
& x_i({k|k}) = x_i(k),\\
& x_i(\kappa+1|k) = A_ix_i(\kappa|k) + B_iu_i(\kappa|k),   \label{MPC_state_dx} \\
& \sum_{ j \in \mathcal{N}^{[s]}}g_{sj} \big(x_{j}(\kappa|k)\big)  \le b_{s}, \\
& u_i(\kappa|k) \le U_i,
\end{align}
\end{subequations}
where $s \in \mathcal{S}_i$ and $\kappa = k, \ldots, k + N - 1$. 

To ensure the stability of distributed MPC without terminal cost, for each $ i \in \mathcal{N}$, we propose to add the following inequality constraints for distributed MPC: 
\begin{subequations}\label{i_terminal}
\begin{align}
& 
\ell_i(x_i (k\! +\! N|k),u_i (k\! +\! N|k)) \le (1\! -\! \alpha) ||x_i (k)||_{Q_i}^2 \! +\! v_i(k), \label{i_terminal_distributed}\\
& \sum_{i\in \mathcal{N}}{v_i(k)} \le 0, 
\end{align}
\end{subequations}
where $v_i(k)$ represents the slack variable for subsystem $i$ at step $k$, and (\ref{i_terminal}) is motivated by the following lemma.

{\bf{Lemma 1.}} For each subsystem $i$ and each step $k$, if there exists an $x_i^*(k+N|k) \in \mathcal{X}_i$ obtained from the optimization problem (\ref{i_MPC}) subject to (\ref{i_constraints}), such that (\ref{i_terminal}) is satisfied, then ${J}\left( x({k}),{\bm{u}}^*({k}) \right) = \sum_{i \in \mathcal{N}}{{J}_i\left( x_i({k}),{\bm{u}}_i^*({k}) \right)}$ is a Lyapunov function for the overall system (\ref{whole_system}),  where 
$\alpha \in (0, 1]$ is the relaxed dynamic programming parameter, and $v_i(k)$ is a slack variable for subsystem $i$. 
\begin{proof}
    Define ${\bm{u}}_i^{*}({k}) = [u_i^{*}(k|k), \ldots, u_i^{*}(k+N -1|k) ]^\intercal$ as the control sequence obtained from (\ref{MPC}) at step $k$, $x_i^*(k+1|k)$ as the state for step $k+1$ obtained at step $k$ with $x_i(k|k)$ and $u_i^{*}(k|k)$, and $ J_i^*(x_i^*({k+1|k}),{\bm{u}}_i^*({k+1}))$ as the optimal cost at step $k+1$ with initial state $x_i^*({k+1|k})$ and optimal input sequence ${\bm{u}}_i^*({k+1})$. 
    Based on the optimality of $ J_i^*(x_i^*({k+1}),{\bm{u}}_i^*({k+1}))$ we have
    \begin{align}\label{feasible_obj}
        J_i(x_i^*({k+1|k}),{\bm{u}}_i^*({k+1})) \le J_i(x_i^*({k+1|k}),{\bm{u}}_i({k+1})),
    \end{align}
    where ${\bm{u}}_i^{*}({k+1}) = [u_i^{*}(k+1|k+1), \ldots, u_i^{*}(k+N|k+1) ]^\intercal$ is the optimal input sequence at step $k+1$, and $ {\bm{u}}_i({k+1}) \allowbreak = \allowbreak [u_i^{*}(k+1|k), \ldots, u_i^{*}(k+N -1|k), \mu_i\big(x^{*}(k+N|k)\big) ]^\intercal$ with $\mu_i\big(x^{*}(k+N|k)\big)$ being a feasible control input as defined in \emph{Theorem 2}. 

    Then, (\ref{i_terminal_distributed}) and the definition of $\ell_i(x_i,u_i)$ implies
    \begin{align*}
     (1-&\alpha)  \ell_i\big(x_i(k),u_i^*(k|k)\big) + v_i(k)  \ge  (1-\alpha) ||x_i(k)||_{Q_i}^2 + \\
     &  + v_i(k) \ge \ell_i\big(x_i^*(k +N|k), \mu\big(x^{*}(k+N|k)\big) \big).
    \end{align*}
    Thus, we get
    \begin{align}
    \ell_i\big(&x_i(k), u_i^*(k|k)\big) + v_i(k) \ge \alpha  \ell_i\big(x_i(k), u_i^*(k|k)\big) + \nonumber \\
    &  + \ell_i\big(x_i^*(k+N|k), \mu\big(x^{*}(k+N|k)\big) \big).  \nonumber
    \end{align}
   Substituting  $\sum\limits_{\kappa = {k+1}}^{{k} + N - 1} {\ell_i(x_{i}^*(\kappa|k),u_{i}^*(\kappa|k))}$ in both sides leads to 
   \begin{align}
   &J_i(x_i({k}),{\bm{u}}_i^*({k})) + v_i(k) \ge \nonumber\\
   & \alpha \ell_i(x_i(k), u_i^*(k|k)) + J_i(x_i^*({k\!+\!1|k}),{\bm{u}}_i({k\!+\!1})) \ge \nonumber\\
   & \alpha \ell_i(x_i(k), u_i^*(k|k)) + J_i(x_i^*({k\!+\!1|k\!+\!1}),{\bm{u}}_i^*({k\!+\!1})).  \label{inequality}
   \end{align}
   Summing both sides of the inequalities in (\ref{inequality}) over $i \in \mathcal{N}$ yields
   \begin{equation}\nonumber
       J(x(\!{k}\!),\!{\bm{u}}^*\!(\!{k}\!)) \! \ge\! \alpha\! \ell(\!x(k),\! u^*\!(k|k)) + J(\!x^*\!({k\!+\!1|k\!+\!1}),\!{\bm{u}}^*\!({k\!+\!1})\!),
   \end{equation}
   which indicate that $J(x^*(\cdot),{\bm{u}}^*(\cdot))$ is a Lyapunov function for the overall system. 
\end{proof}
In addition, if \emph{Lemma 1} holds, following \emph{Theorem 6.20} in \cite{grune2017nonlinear}, the performance of the distributed MPC controller relates to the infinite-horizon controller as
\begin{equation}
J_{\infty}^{\mathrm{cl}}\big(x(k)\big) \le J\big(x(k), {\bm{u}}^*(k) \big)/\alpha \le V_{\infty}\big(x(k)\big)/\alpha ,
\end{equation}
where $J_{\infty}^{\mathrm{cl}}\big(x(k)\big) := \sum_{\kappa=k}^{\infty}\ell(x^*(\kappa|\kappa),u^*(\kappa|\kappa))$ represents the cost of applying MPC in closed-loop up to $k \rightarrow \infty$, and $V_{\infty}\big(x(k)\big):= \min \sum_{\kappa=k}^{\infty}\ell(x(\kappa|k),u(\kappa|k))$ is the cost of the infinite horizon optimal controller, respectively. 

Therefore, the optimization problem to be solved for each subsystem $i$ in cooperative distributed MPC setting becomes 
\begin{equation}\label{i_MPC_new}
\mathop {\min }\limits_{\scriptstyle{\bm{u}}_i({k}), v_i(k)}\ {J_i}\left( x_i({k}),{\bm{u}}_i({k}) \right)  := \sum\limits_{\kappa = {k}}^{{k} + N - 1} {\ell_i(x_{i}(\kappa|k),u_{i}(\kappa|k))},
\end{equation}
subject to
\begin{subequations}\label{i_final_constraints}
\begin{align}
&x_i({k|k}) = x_i(k),\\
& x_i(\kappa+1|k) = A_ix_i(\kappa|k) + B_iu_i(\kappa|k),  \\
& \sum_{ j \in \mathcal{N}^{[s]}}g_{sj} \big(x_{j}(\kappa|k)\big) \le  b_{s}, \\
& u_i(\kappa|k) \le U_i,\\
& \ell_i(x_i (k\! +\! N|k),u_i (k\! +\! N|k)) \! \le \! (\!1\! - \! \alpha\!) ||\!x_i (k)\!||_{Q_i}^2 \!+\! v_i(k), \label{convex_cons}\\
& \sum_{i\in \mathcal{N}}{v_i(k)} \le 0, \label{slack}\\
&s \in \mathcal{S}_i, \ \kappa = k, \ldots, k + N. \nonumber
\end{align}
\end{subequations}


{\bf{Lemma 2.}} If the system $ x_i(k+1) =A_ix_i(k)+ B_iu_i(k)$ is controllable in feasible set $\mathcal{X}_i$ with initial state $x_i(k) \in \mathcal{X}_i$ and prediction horizon $N$, the solution of optimization problem (\ref{i_MPC_new}) subject to (\ref{i_final_constraints}) always exists. Furthermore, the resulting problem is a convex optimization problem.
\begin{proof}
    Given the system  $ x_i(k+1) =A_ix_i(k)+ B_iu_i(k)$ is controllable for prediction horizon $N$, we can find a control sequence  $u_i(\kappa|k), \kappa = k, \ldots, k + N $ and $v_i(k) = 0$ such that $x_i(\kappa|k) $ satisfy constraints in (\ref{i_final_constraints}), which means a feasible solution can be found. 
     As $x_i(k)$ is known, (\ref{convex_cons}) defines a convex constraints. Then, the original optimization remains convex while adding (\ref{slack}). 
\end{proof}


{\bf{Remark 1.}} If  $v_i(k) = 0, \ i \in \mathcal{N}$, then ${J}_i\left( x_i({k}),{\bm{u}}_i^*({k}) \right)$ from \emph{Lemma 1} is a Lyapunov function of subsystem $i$, and the resulting ${J}\left( x({k}),{\bm{u}}^*({k}) \right)$ is still a Lyapunov function for the system, but constraint (\ref{i_terminal_distributed}) becomes more conservative.

{\bf{Remark 2.}} If  $v_i(k) = 0, \ i \in \mathcal{N}$ and $\alpha = 1$, then (\ref{i_terminal}) is reduced to a stability condition with a terminal equality constraint, i.e., $x_i (k + N) = \bm{0}$.

\section{Distributed synthesis of cooperative distributed MPC}\label{section_distributed}
In Section~\ref{section_stability}, the general optimization problem for each agent with stability guarantee for cooperative distributed MPC has been formulated. 
This section extends the approach in \cite{liu2024achieving} to distributed optimization with convex local and coupled constraints, and applies it to achieve the distributed synthesis of distributed MPC. 

For compactness, we stack all decision variables into one vector $y_i(k) \in \mathbb{R}^{(n_i+m_i)(N+1) + 1}$ as
\begin{equation}\nonumber
    y_i(k) = [x_i^\intercal({k}), \cdots,  x_i^\intercal({k+N}), u_i^\intercal({k}), \cdots,  u_i^\intercal({k+N}), v_i({k})]^\intercal.
\end{equation} 
Then, the MPC optimization problem for the overall system can be written compactly as 
\begin{subequations}\label{overall}
\begin{align}
 \mathop {\min }\limits_{\scriptstyle y(k)} \ &  \sum\limits_{i \in \mathcal{N}} {\frac{1}{2}y_i^\intercal(k) H_i y_i(k)},   \\
 \mathrm{s.t.}\ 
& F_i y_i(k) = 0,  \label{collect_state}\\
& \sum_{j\in \mathcal{N}^{[s]}} c_{sj}\big(y_j(k)\big) \le d_{s}, \label{coupled} \\
& s \in \mathcal{S}, \ i \in \mathcal{N},  \nonumber
\end{align}    
\end{subequations}
where (\ref{collect_state}) denotes the compact version of state function at step $k$, $H_i$ and $F_i$ are matrices of appropriate dimensions, $y(k) = \mathrm{col}_{i \in \mathcal{N}}\big(y_i(k)\big)$, $\mathcal{S}$ is the set of coupled constraints; $c_{sj}(\cdot)$, $j \in \mathcal{N}^{[s]}$ represents the convex function associated to inequality constraint $s$ and $d_s$ is the corresponding constant.  Then, (\ref{coupled}) represents all the inequality constraints, and if $ |\mathcal{N}^{[s]}|  = 1 $, (\ref{coupled}) represents a local constraint.  For the sake of simplicity, we still use $s$ as the index of the inequality constraints in the following. 

For each coupled constraint $s$, we define a doubly stochastic 
weight matrix $P^{[s]} = [p_{ij}^{[s]}] \in \mathbb{R}^{|\mathcal{N}^{[s]}|\times|\mathcal{N}^{[s]}|}$ for agents to weigh the information exchanged with their neighbors:
\begin{equation}\label{weight_matrix}
p_{ij}^{[s]} 
\begin{cases}
    > 0, & \mathrm{if} \ j \in \mathcal{N}_i^{[s]},\\
    = 0, & \mathrm{otherwise},
\end{cases}, \sum_{j \in \mathcal{N}}p_{ij}^{[s]} = 1, \ \forall i \in \mathcal{N}. 
\end{equation}

Let us introduce $z_{j}^{[s]}$ as the slack variable of subsystem $j$ for coupled constraint $s$, and define $\bm{c}_s (y) = \mathrm{col}_{j \in \mathcal{N}^{[s]}}\big(c_{sj}(y_j)\big)$, $z^{[s]} = \mathrm{col}_{j \in \mathcal{N}^{[s]}}\big(z_{j}^{[s]}\big)$,  $z = \mathrm{col}_{s \in \mathcal{S}}(z^{[s]})$, and $\bm{1}$ as a vector of all ones. Then, we have the following problem 
\begin{subequations}\label{overall_decomposed}
\begin{align}
 \mathop {\min }\limits_{\scriptstyle y(k)} \ &  \sum\limits_{i \in \mathcal{N}} {\frac{1}{2}y_i^\intercal(k) H_i y_i(k)},  \\
 \mathrm{s.t.}\ 
& F_i y_i(k) = 0, \\
& \bm{c}_s\big(y(k)\big) + (I - P^{[s]})z^{[s]}(k) \le \frac{\bm{1}}{|\mathcal{N}^{[s]}|} d_{s}, \label{overall_decomposed_decouple} \\ 
& s \in \mathcal{S}, \nonumber
\end{align}    
\end{subequations} 
where the coupled constraint $s$ is decoupled into $|\mathcal{N}^{[s]}|$ constraints by introducing the slack variable $z^{[s]}$, and the constant $d_s$ is distributed equally among the subsystems corresponding to inequality constraint $s$. 

{\bf{Proposition 1.}}
The problems in (\ref{overall}) and (\ref{overall_decomposed}) share the same objective function and are equivalent in the sense that 
\begin{enumerate}[i)]
    \item for any feasible solution $\big(y(k), z(k)\big)$ to (\ref{overall_decomposed}), $y(k)$ is feasible for (\ref{overall}),
    \item for any feasible solution $y(k)$ to (\ref{overall}), there exist $z(k)$ such that $\big(y(k), z(k)\big)$ is feasible for (\ref{overall_decomposed}).
\end{enumerate}
\begin{proof}
The proof is a direct extension of \emph{Proposition 1} in  \cite{liu2024achieving}.



\end{proof}

Then, (\ref{overall_decomposed}) can be decomposed among each agent $i$ as
\begin{subequations}\label{sub_decomposed}
\begin{align}
 \mathop {\min }\limits_{\scriptstyle y_i(k)} \ & \frac{1}{2}y_i^\intercal(k) H_i y_i(k),  \\
 \mathrm{s.t.}\ 
& F_i y_i(k) = 0,\\
& c_{si}\big(y_i(k)\big) + z_{i}^{[s]}(k) -  \sum_{j \in \! \mathcal{N}^{[s]}} p_{ij}^{[s]}z_j^{[s]}(k) \le \frac{1}{|\mathcal{N}^{[s]}|} d_{s},\label{const_relax} \\ 
& s \in \mathcal{S}_i. \nonumber
\end{align}    
\end{subequations} 

As indicated by \cite{liu2024achieving}, problem (\ref{sub_decomposed}) may become infeasible for some $z^{[s]}, s \in \mathcal{S}_i$ even with all constraints are linear. In the following, we introduce a constraint tightening parameter $\delta_{s} \in (0,1]$ and a relaxed variable $\rho_i$ for each constraint $s$ with $0 \le \rho_i \le \delta_{s}$ to ensure the feasibility at each iteration, which is analyzed in \emph{Theorem 3}.  Let us define the minimum value of the objective in (\ref{sub_decomposed}) as a function of $z_i(k)$, 
so that problem (\ref{sub_decomposed}) becomes
\begin{subequations}\label{sub_decomposed_tightening}
\small
\begin{align}
 \phi_i\big(z_i&(k)\big) :=  \mathop {\min }\limits_{\scriptstyle y_i(k) \hfill}   \frac{1}{2}y_i^\intercal(k) H_i y_i(k) + w \sum_{s\in \mathcal{S}_i}\rho_{si} ,  \\
 \mathrm{s.t.} \ 
& F_i y_i(k) = 0,\\
& c_{si}\big(y_i(k)\big) \! +\! z_{i}^{[s]}(k) \!- \! \sum_{j \in \! \mathcal{N}^{[s]}} \!p_{ij}^{[s]}z_j^{[s]}(k) \le \frac{1\! -\! \delta_{s} \!+\! \rho_{si} }{|\mathcal{N}^{[s]}|} d_{s}, \label{const_relax_tightening} \\ 
& 0 \le \rho_{si} \le \delta_{s}, \\
& s \in \mathcal{S}_i, \nonumber
\end{align}    
\end{subequations} 
where $w$ is a large positive scalar. 

The Lagrangian of problem (\ref{sub_decomposed_tightening}) is
{\scriptsize
\begin{align}
    &L_i\big(y_i(k), \mu_i, \lambda_i \big) = \frac{1}{2}y_i^\intercal(k) H_i y_i(k) + w \sum_{s\in \mathcal{S}_i}\rho_{si} + \Big\langle \mu_i, \  F_i y_i(k) \Big\rangle \nonumber \\
    &\!+\! \sum_{s \in \mathcal{S}_i}\!\Big\langle \!\lambda_i^{[s]}, \!c_{si}\big(\!y_i(k)\!\big) \! +\! z_{i}^{[s]}(k) \!- \! \sum_{j \in \! \mathcal{N}^{[s]}} \!p_{ij}^{[s]}z_j^{[s]}\!(k) \!- \! \frac{1\! -\! \delta_{s} \!+\! \rho_{si} }{|\mathcal{N}^{[s]}|} d_{s} \!\Big\rangle, \label{Lagrange}
\end{align}}
\par where $\mu_i$ and $\lambda_i = \mathrm{col}_{s \in \mathcal{S}_i}\big(\lambda_i^{[s]} \big) $ are the multipliers of problem (\ref{sub_decomposed_tightening}) associated them with equality and inequality constraints, respectively.

{\bf{Theorem 3.}} 
If for any $ s \in \mathcal{S}_i,  \ i \in \mathcal{N}$, 
the scalar $z_{i}^{[s]}(k)\in \big[-\frac{\delta_{s} }{2\cdot|\mathcal{N}^{[s]}|} d_{s}, \frac{\delta_{s} }{2\cdot|\mathcal{N}^{[s]}|} d_{s}\big] \subset \mathbb{R}$, then, the following hold
\begin{enumerate}[i)]
    \item a feasible solution $y_i(k)$ for (\ref{sub_decomposed_tightening}) can be found,
    \item the resulting $y_i(k)$ obtained by solving (\ref{sub_decomposed_tightening}) is feasible for problem (\ref{sub_decomposed}),
    \item a feasible solution $y(k)$ for (\ref{overall}) can be obtained by concatenating the solution of (\ref{sub_decomposed_tightening}) solved locally for each $i$. 
\end{enumerate}
\begin{proof}
i) We first define a projection operator to ensure the existence of $z_{i}^{[s]}(k)\in \big[-\frac{\delta_{s} }{2\cdot|\mathcal{N}^{[s]}|} d_{s}, \frac{\delta_{s} }{2\cdot|\mathcal{N}^{[s]}|} d_{s}\big]$, for that, each $z_i^{[s]}(k)$ is replaced by
\begin{equation}
    z_i^{[s]}(k) \leftarrow \mathrm{Proj}_{\frac{\delta_{s} d_{s} }{2\cdot|\mathcal{N}^{[s]}|} [-1, 1]}\big(z_i^{[s]}(k) \big),
\end{equation}
where $\mathrm{Proj}_{[-a, a]}(\cdot)$ represents the projection onto the set $[-a, a]$. 


Having (\ref{weight_matrix}) and given that $z_{i}^{[s]}(k)\in \big[-\frac{\delta_{s} }{2\cdot|\mathcal{N}^{[s]}|} d_{s}, \frac{\delta_{s} }{2\cdot|\mathcal{N}^{[s]}|} d_{s}\big]$, for $\rho_{si} = \delta_{s}$ we have
\begin{equation}\nonumber
    z_{i}^{[s]}(k) \!- \! \sum_{j \in \! \mathcal{N}^{[s]}} \!p_{ij}^{[s]}z_j^{[s]}(k) \le  \frac{\delta_{s} }{|\mathcal{N}^{[s]}|} d_{s} = \frac{\rho_{si} }{|\mathcal{N}^{[s]}|} d_{s}.
\end{equation}
Moreover, since for any $s\in \mathcal{S}_i$, according to \emph{Assumption~1}, we have $\{ g_{si} (x_{i}) \le 0 , s \in \mathcal{S}_i \} \neq \varnothing$,  a $y_i(k)$ can be found such that
\begin{equation}\nonumber
c_{si} (y_{i}(k)) \le 0 \le   \frac{1\! -\! \delta_{s}}{|\mathcal{N}^{[s]}|} d_{s},
\end{equation}
for any $s \in \mathcal{S}_i$. Thus, a feasible solution for (\ref{sub_decomposed_tightening}) can be found. 

ii) According to i), a feasible $y_i(k)$ and $\rho_{si} \in [0, \delta_{s}]$ can be found by solving (\ref{sub_decomposed_tightening}). Then, $\frac{1 - \delta_{s} + \rho_{si} }{|\mathcal{N}^{[s]}|} d_{s} \le \frac{1}{|\mathcal{N}^{[s]}|} d_{s} $, and thus a feasible solution $y_i(k)$ for (\ref{sub_decomposed_tightening}) is also feasible for (\ref{sub_decomposed}).

iii) The statement in iii) holds trivially according to i), ii), and \emph{Proposition~1}.    
\end{proof}

Note that in (\ref{sub_decomposed_tightening}), each subsystem solves its local optimization problem at each iteration with equally distributed constant $\frac{1\! -\! \delta_{s} \!+\! \rho_{si} }{|\mathcal{N}^{[s]}|} d_{s}$. However, distributing constant $(1\! -\! \delta_{s} \!+\! \rho_{si} )d_{s}$ equally is conservative, and an adaption is developed for distributed optimization during iterations. Let us define a gap for constraint $s$ with subsystem $i$ after each iteration as 
\begin{equation}\label{delta_gap}
    \Delta d_{si} = \frac{1\! -\! \delta_{s}\! + \! \rho_{si} }{|\mathcal{N}^{[s]}|} d_{s}\! -\! \Big(\! c_{si}\!\big(y_i(k)\!\big) \! +\! z_{i}^{[s]}\!(k) \!- \! \sum_{j\! \in \! \mathcal{N}^{\![s]}} \!p_{ij}^{[s]}z_j^{[s]}\!(k) \!\Big),
\end{equation}
Then, the distribution of the coupled constraint among different agents can be updated as
\begin{align}
    &  c_{si} (y_{i}(k)) \leftarrow c_{si} (y_{i}(k)) + \Delta d_{si}, \label{update_distribution_1}\\
    &  d_s \leftarrow d_s + \sum_{i \in \mathcal{N}^{[s]}}\Delta d_{si}.\label{update_distribution_2}
\end{align}

The feasibility and convergence analysis after the updating with (\ref{delta_gap})-(\ref{update_distribution_2}) is given in \emph{Theorem 4} below.

Having (\ref{sub_decomposed_tightening}), the optimization problem for the overall system with constraint tightening can be formulated as 
\begin{equation}\label{phi}
    \min_{z(k)} \Big\{ \phi\big(z(k)\big) = \sum_{i \in \mathcal{N}}\phi_i\big(z_i(k)\big)\Big\}.
\end{equation}
Then, a solution to the original problem (\ref{overall}) can be obtained by substituting $z(k)$ into (\ref{sub_decomposed_tightening}) and solving the resulting local optimization problem for each agent $i$ in parallel.

{\bf{Lemma 3}} \citep{liu2024achieving}. In (\ref{phi}), $\phi\big(\cdot\big)$ is convex and differentiable and the gradients of $\phi\big(\cdot\big)$ can be computed as
\begin{equation}\label{gradient}
    \nabla_{z^{[s]}(k)} \phi\big(z(k)\big) = \big(I - P^{[s]} \big)^\intercal \mathrm{col}_{s \in \mathcal{S}_i}\big(\lambda_i^{[s]} \big). 
\end{equation}

Then, following \emph{Theorem 3} and \emph{Lemma 3}, $z_i^{[s]}(k)$ can be updated by
\begin{equation}\label{projection}
    z_i^{[s]}(k) \leftarrow \mathrm{Proj}_{\frac{\delta_{s} d_{s} }{2\cdot|\mathcal{N}^{[s]}|} [-1, 1]}\Big(z_i^{[s]}(k) - \gamma\nabla_{z_i^{[s]}(k)} \phi\big(z(k)\big) \Big),
\end{equation}
where $\gamma$ is the step size of updating $z_i^{[s]}(k)$ during iterations.

Based on \emph{Theorem 3}, we construct Algorithm~\ref{alg_cdmpc} to obtain the solution for each local agent $i$ in parallel. In this algorithm, $q$ is the iteration index, $q_\mathrm{max}$ is the maximum number of iterations, $\Delta d_{s} = [\Delta d_{s1},\cdots, \Delta d_{s|\mathcal{N}^{[s]}|}]^\intercal$, and $d_\mathrm{min}$ represents a small value for iteration. 

\begin{algorithm}[h]
  \caption{Violation-free distributed optimization method}  
  \label{alg_cdmpc}
  \begin{algorithmic}[1]  
    \Require  arbitrary variable $z(k)$, step size $\gamma$, $q_{\mathrm{max}}$, $d_\mathrm{min}$, and $\delta_s, \ s \in \mathcal{S}$
    \Ensure  $y_i\big(z_i(k)\big), \ i \in \mathcal{N}$
    \For{$q \in 1, \cdots, q_{\mathrm{max}}$}
      \For {$i \in \mathcal{N}$}
      \State collect $z_j^{[s]}(k)$ for each $j \in \mathcal{N}_i^{[s]}$, $\forall s \in \mathcal{S}_i$
      \State compute $y_i\big(z_i(k)\big)$, $\rho_{si}$, the multipliers $\mu_i\big(z_i(k)\big)$, and $\lambda_i\big(z_i(k)\big)$ via (\ref{sub_decomposed_tightening})
      \State collect $\lambda_j^{[s]}\big(z_j(k)\big)$ for each $j \in \mathcal{N}_i^{[s]}$, $\forall s \in \mathcal{S}_i$
      \State compute $\Delta d_{si}$ via (\ref{delta_gap})
      \If{$||\Delta d_{s}|| > d_\mathrm{min}$, $ s \in \mathcal{S}_i$}
      \State update $c_{si} (y_{i}(k))$ by (\ref{update_distribution_1}) and $d_{s}$ by (\ref{update_distribution_2}) 
    \Else 
        \State compute $\nabla_{z_i^{[s]}(k)} \phi\big(z(k)\big)$, $\forall s \in \mathcal{S}_i$ via (\ref{gradient})
        \State update $z_i^{[s]}(k)$ by (\ref{projection})
        \EndIf
      \EndFor
      \EndFor
  \end{algorithmic}  
\end{algorithm}



{\bf{Theorem 4.}} Given that \emph{Lemma 3} holds, and defining 
$z^{(0)}(k)$ as the initial value of $z(k)$, $z^*(k)$ as the optimal value of $z(k)$, the solution to the problem in (\ref{sub_decomposed_tightening}) obtained by Algorithm 1 is feasible for (\ref{overall}), and the overall objective function obtained by solving the problem in (\ref{sub_decomposed_tightening}) satisfies $\phi^{(q+1)}\big(z(k)\big) \le \phi^{(q)}\big(z(k)\big)$ for all $ q$. 
\begin{proof}
   Algorithm 1 consists of two parts defined by the ``if-else'' statement. We will demonstrate the convergence of the solution to (\ref{sub_decomposed_tightening}) for both branches of the ``if-else'' statement. 

   i) For the case $||\Delta d_{s}|| > d_\mathrm{min}$, $ s \in \mathcal{S}_i$, the proof is based on induction. An initial feasible solution to (\ref{sub_decomposed_tightening}) can be found directly by substituting a feasible $z_i^{[s]}(k)$ from (\ref{projection}) in (\ref{sub_decomposed_tightening}).
   
   Let us define a solution $y_i^{(q)}(k)$ to (\ref{sub_decomposed_tightening}) at iteration $q$ with optimal objective function value as $\phi_i^{(q)}\big(z_i(k)\big)$.
   
   After updating $c_{si} (y_{i}(k))$ with (\ref{update_distribution_1}) and $d_{s}$ with (\ref{update_distribution_2}), we can observe that $y_i^{(q)}(k)$ is still feasible in the updated optimization problem (\ref{sub_decomposed_tightening}). \\
   Then, as we have a newly defined optimization problem (\ref{sub_decomposed_tightening}) for iteration $q+1$ with updated constraint in (\ref{const_relax_tightening}). Since we have more relaxed constraints in $ \phi_i^{(q+1)}\big(z(k)\big)$,  we have
   \begin{equation}
        \phi_i^{(q+1)}\big(z(k)\big) \le \phi_i^{(q)}\big(z(k)\big),
   \end{equation}
   which indicates that for the overall objective function it holds that $\phi^{(q+1)}\big(z(k)\big) = \sum_{i\in \mathcal{N}}\phi_i^{(q+1)}\big(z(k)\big) \le \phi^{(q)}\big(z(k)\big)$. 
   
    ii) For the case $||\Delta d_{s}|| \le d_\mathrm{min}$, $\forall s \in \mathcal{S}_i$: 
     let us define $\theta_{q} = \nabla_{z^{(q)}(k)} \phi\big(z(k)\big)$. Then, the projected  $z^{(q+1)}$ is given by
    \begin{equation}
        z^{(q+1)} = \mathrm{Proj}_{\frac{\delta_{s} d_{s} }{2\cdot|\mathcal{N}^{[s]}|} [-1, 1]}\Big(z^{(q)}(k) - \gamma\theta_{q} \Big),
    \end{equation}
    where $\gamma$ represents the constant step size.
   Let $\chi^{(q+1)} = z^{(q)}(k) - \gamma \theta^{(q)}$. Then, we have\vspace{-\baselineskip}
   {\small
   \begin{align*}
          &||\chi^{(q+1)}(k) - z^*(k)||_2^2 \\
        & = ||z^{(q)}(k) - \gamma\theta_{q} - z^*(k)||_2^2\\
        & \le ||z^{(q)}\!(k)- z^*(k)||_2^2  -  2  \gamma \theta_q^\intercal\big(z^{(q)}(k)- z^*(k)\big)+ \gamma^2 ||\theta_{q}||_2^2,\\
        & \le ||z^{(q)}\!(k)\!- \!z^*\!(k)||_2^2  -  2 \gamma \Big(\!\phi\big(\!z^{(q)}\!(k)\!\big) \!-\! \phi\big(z^*\!(k)\!\big)\! \Big)+ \gamma^2 ||\theta_{q}||_2^2,
    \end{align*}}
    \par where the last inequality is the result of convex function $\phi(\cdot)$. Therefore, we can write
    which indicates that \vspace{-\baselineskip}
    {\small\begin{align*}
        &||z^{(q+1)}(k) - z^*(k)||_2^2 \le ||\chi^{(q+1)}(k) - z^*(k)||_2^2\\
      & \le \! ||z^{(q)}\!(k) \!-\! z^*\!(k)||_2^2  -  2 \gamma \Big(\!\phi\big(z^{(q)}\!(k)\big)\! -\! \phi\big(z^*\!(k)\big)\! \Big)+ \gamma^2 ||\theta_{q}||_2^2.
    \end{align*}}
\par Then, following the direct results of convergence analysis of the subgradient method, we have $\chi^{(q+1)}(k)$ converge to the optimal value $ z^*(k) $ \citep{boyd2003subgradient}, and they are omitted here for brevity. 

According to \emph{Theorem~3}, each $z(k)$ corresponds to a feasible solution $y(k)$ for problem (\ref{sub_decomposed_tightening}), i.e., for each $z(k)$ a feasible solution $y(k)$ can be found by solving problem (\ref{sub_decomposed_tightening}), i.e., $\phi^{(q+1)}\big(z(k)\big) \le \phi^{(q)}\big(z(k)\big)$ can be ensured for all $ q$.
\end{proof}

\section{Numerical example}\label{section_numerical}
In the case study, we consider a platoon of autonomous vehicles, as described in \cite{he2024approximate}, where three follower vehicles follow a leader vehicle on a straight highway. The leader has a constant speed $v_\mathrm{ref} = 20$~m/s. The objective is to control the followers to keep a desired distance $d = 40$~m with respect to the predecessor vehicle with the desired speed $v_\mathrm{ref} = 20$~m/s, while ensuring safety. 

\begin{figure}[htb]
\begin{center}
\includegraphics[width=8cm]{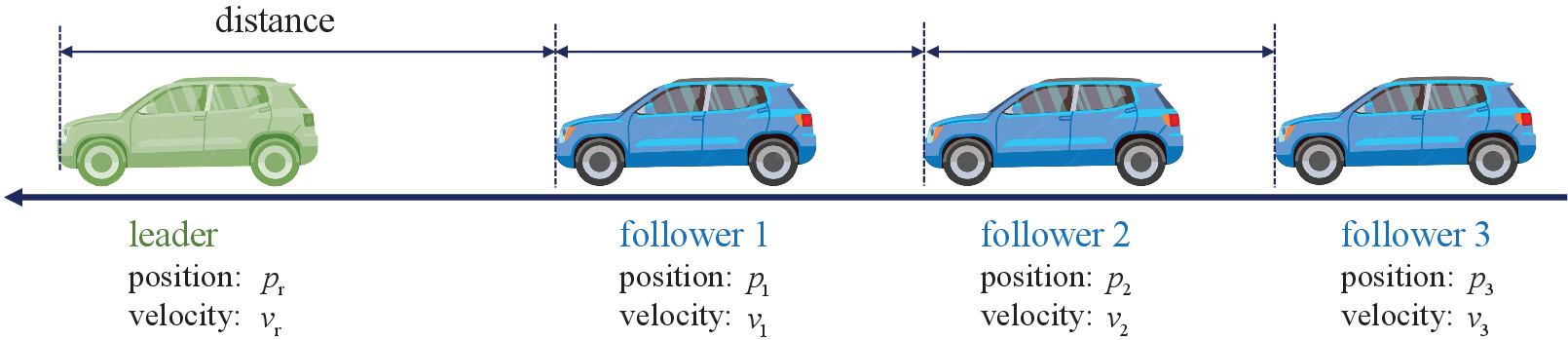}    
\caption{Distributed MPC for connected and automated vehicles.}
\label{smart_car}
\end{center}
\end{figure}

We consider the model of vehicle $i$ as $s_i(k+1) = s_i(k) + v_i(k)T$, $v_i(k+1) = v_i(k) + \frac{bT}{m}u_i(k) - cTv_i^2(k) - \mu mg$, where $s_i(k)$, $v_i(k)$, and $u_i(k)$ are the position, speed, and input force of vehicle $i$ at time step $k$, $T = 0.1$~s is the sampling time, $b= 3700$~N, $c = 0.5$~kg/m, $\mu = 0.01 $, $m = 1000$~kg, $g = 10$~m/s$^2$. The speed limit is $v_\mathrm{lim} = 25$~m/s. 
This example considers the cruising control problem with desired speed $v_\mathrm{ref} = 20$~m/s. Similar with \cite{he2024approximate}, we linearize $V_i(v) = cv_i^2$ around the desired speed as $V_i(v) = c_{\mathrm{app}}v_i$ with $c_{\mathrm{app}} = 13cv_\mathrm{lim}/8$. In this context, the model of cars considered in this paper can also be written as $s_i(k+1) = s_i(k) + v_i(k)T$, $v_i(k+1) = v_i(k) + \frac{bT}{m}u_i(k) - c_{\mathrm{app}}Tv_i(k) - \mu mg$.

The constraints considered include limitations on the distance between two adjacent vehicles, the velocity, and the traction/brake force input. Specifically, the distance between vehicles is restricted by $|s_{i}(k) - s_{i-1}(k)| \ge 10$ m, the velocity by 5 m/s $\le v_i(k) \le$ 25 m/s, and the control input by $|u_i(k)| \le 1 $, for $i = 1, 2, 3$ and where $s_0(k)$ represents the position of the leader vehicle.  
The stage cost is chosen as $\ell_i (x_i(k), u_i(k)) = ||x_i(k) - x_{\mathrm{r},i} ||_{Q}^2 + ||u_i(k) - u_{\mathrm{r},i} ||_{R}^2$, with $x_i(k) = [v_i(k), s_i(k)]^\intercal$, $x_{\mathrm{r},i} = [v_\mathrm{ref}, \quad s_0(k) - i\cdot d]^\intercal$, $u_{\mathrm{r},i} = u_{0}$, $Q = \mathrm{diag}([1\quad 0.5])$, and $R = 0.1$.

\begin{figure}[htb]
\begin{center}
\includegraphics[width=8cm]{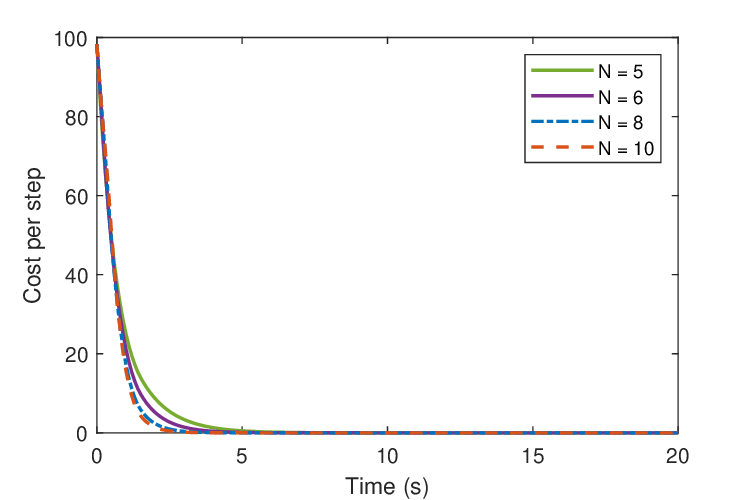}    
\caption{Cost per step for different horizons.}
\label{cost per step}
\end{center}
\end{figure}

\begin{figure}[htb]
\begin{center}
\includegraphics[width=8cm]{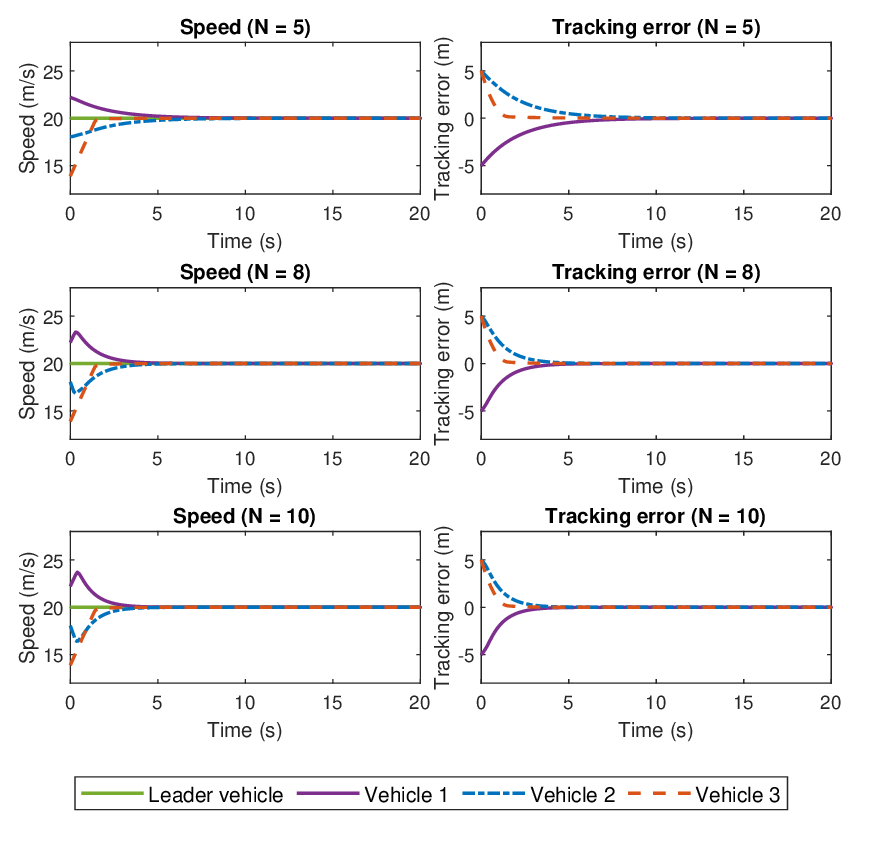}    
\caption{Speed and position tracking of each vehicle.}
\label{tracking}
\end{center}
\end{figure}

We perform simulations with the deterministic case for 200 control steps with $\delta = 0.5$ and $q_\mathrm{max} = 5$. The simulation results are shown in Fig.~\ref{cost per step}, Fig.~\ref{tracking}, and Table~\ref{dmpc-closed-loop}. All results are obtained in MATLAB (R2019b) on a desktop with an Intel Xeon W-2223 CPU and 8GB RAM. The stability of the overall system is reflected in Fig.~\ref{cost per step}, where the cost per step is shown, which finally converges to zero.  The speed and position tracking results are illustrated in Fig.~\ref{tracking}, indicating the stability of the overall system.

\begin{table}[h]
\footnotesize
\caption{Simulation results of distributed MPC}\label{dmpc-closed-loop}
\begin{tabular}{ccccc} \hline
\multirow{2}*{Horizon}  & RDP parameter  & \multicolumn{2}{c}{ Solution time (s)}& \multirow{2}*{Total cost}\\ 
& $\alpha$ &  mean & max&  \\ \hline
\multirow{4}*{N = 5} & 0.10&  0.38 &  0.52 &  824.4694\\
 & 0.30&  0.37 &  0.55 &  824.4694\\
 & 0.50&  - &  - &  - \\
 & 0.70&  - &  - &  -\\ \hline
\multirow{4}*{N = 8} & 0.10&  0.49 &  1.31 &  656.2502\\
& 0.30&  0.50 &  1.38 &  656.2502\\
 & 0.50&  0.51 &  1.25 &  656.2502\\
 & 0.70&  - &  - &  -\\\hline
\multirow{4}*{N = 10} & 0.10&  0.63 &  2.98 &  640.3915\\
& 0.30&  0.60 &  4.23 &  640.3915\\
& 0.50&  0.63 &  2.98 &  640.3915\\
& 0.70&  0.62 &  6.46 &  640.3915\\ 
\hline
\end{tabular}
\end{table}

It can be observed from Table~\ref{dmpc-closed-loop} that a larger horizon leads to better performance in terms of total cost while the solution time increases. 
A larger $N$ and a smaller $\alpha$ yield a larger region of attraction, which is reflected in Table~\ref{dmpc-closed-loop} where a feasible solution cannot be found with $N = 5$, $\alpha = 0.50$, or $ \alpha = 0.70$ and $N = 8$, $\alpha = 0.70$. 
Combined with Fig.~\ref{cost per step} and Fig.~\ref{tracking}, it can be observed that the stability of the overall system is ensured as long as the problem corresponding to the given $N$ and $\alpha$ is feasible. The simulation results further demonstrate that the proposed distributed MPC approach guarantees stability, with each agent computing its local control input in parallel, provided that feasible values of $N$ and $\alpha$ are selected. 

\section{Conclusions}\label{conclusions}
In this paper, a distributed model predictive control (MPC) formulation without terminal cost has been developed, where an explicit stability condition, based on relaxed dynamic programming, is included as an additional constraint.  A distributed synthesis approach has been developed to decouple the resulting problem, enabling each agent to solve its local optimization problem in parallel.  
In addition, a violation-free distributed optimization approach has been developed with constraint tightening to ensure the feasibility throughout the iterations.
A numerical example has been presented to demonstrate that the developed distributed MPC approach ensures closed-loop stability with each feasible control sequence while allowing each agent to compute its control input in parallel.

In the future, we will extend the framework to nonlinear systems, thereby improving its applicability to a broader range of scenarios. Furthermore, the development of robust and stochastic distributed MPC approaches offers a promising direction for handling uncertainties. 

\section*{Acknowledgements}
This work is supported by the European Research Council (ERC) under the European Union’s Horizon 2020 research and innovation programme (Grant agreement No.~101018826 - CLariNet). The work of the first author is also supported by China Scholarship Council (No.~202007090003). 





\bibliographystyle{elsarticle-harv}
\bibliography{autosam}           



\end{document}